\documentclass[aps,prd,twocolumn,showpacs,eqsecnum,floats,nofootinbib]{revtex4}
\usepackage{graphicx}
\usepackage{amsfonts}
\usepackage{bm}

\def\d#1{\partial_{#1}}

\begin{document}
\title{3D simulations of linearized scalar fields in Kerr spacetime}
\author{Mark A. Scheel${}^1$, Adrienne L. Erickcek${}^2$,
Lior M. Burko${}^3$, Lawrence E. Kidder${}^4$, Harald P. Pfeiffer${}^4$, 
and Saul A. Teukolsky${}^4$}

\affiliation{${}^1$ Theoretical Astrophysics 130-33, California
Institute of Technology, Pasadena, California 91125} 
\affiliation{${}^2$ Department of Physics,
		    Princeton University, Princeton, New Jersey 08544} 
\affiliation{${}^3$ Department of Physics, University of Utah,
	Salt Lake City, Utah 84112}
\affiliation{${}^4$ Center for Radiophysics and Space Research,
Cornell University, Ithaca, New York 14853}

\date{\today}

\begin{abstract}
We investigate the behavior of a dynamical scalar field on a fixed
Kerr background in Kerr-Schild coordinates
using a 3+1 dimensional spectral evolution code, and
we measure the power-law tail decay that occurs at late times.  We
compare evolutions of initial data proportional to 
$f(r) Y_{\ell m}(\theta,\phi)$, where $Y_{\ell m}$ is a spherical
harmonic and $(r,\theta,\phi)$ are
Kerr-Schild coordinates, to that of initial data proportional to
$f(r_{\rm BL}) Y_{\ell m}(\theta_{\rm BL},\phi)$, where 
$(r_{\rm BL},\theta_{\rm BL})$ are
Boyer-Lindquist coordinates. We find that although these two cases are
initially almost identical, the evolution can be quite different at
intermediate times; however, at late times the power-law decay rates
are equal.
\end{abstract}

\maketitle

\section{Introduction}

The propagation of classical scalar fields in a fixed black hole
spacetime has been studied extensively ever since the work of
Price~\cite{Price1972} that described the behavior of such fields in
the Schwarzschild geometry.  Higher-spin fields, such as linearized
gravitational perturbations, behave qualitatively similar to the
zero-spin case, and therefore scalar fields are often used to gain
insight into more general situations.  Although the behavior of scalar
fields in Schwarzschild spacetime is well-understood, the situation
for a Kerr background geometry is still under active investigation and
has been a topic of some controversy (see, {\it
e.g.,\/}~\cite{Poisson2002,Burko2002} and references therein).

\begin{figure} 
\begin{center}
\includegraphics[width=3.0in]{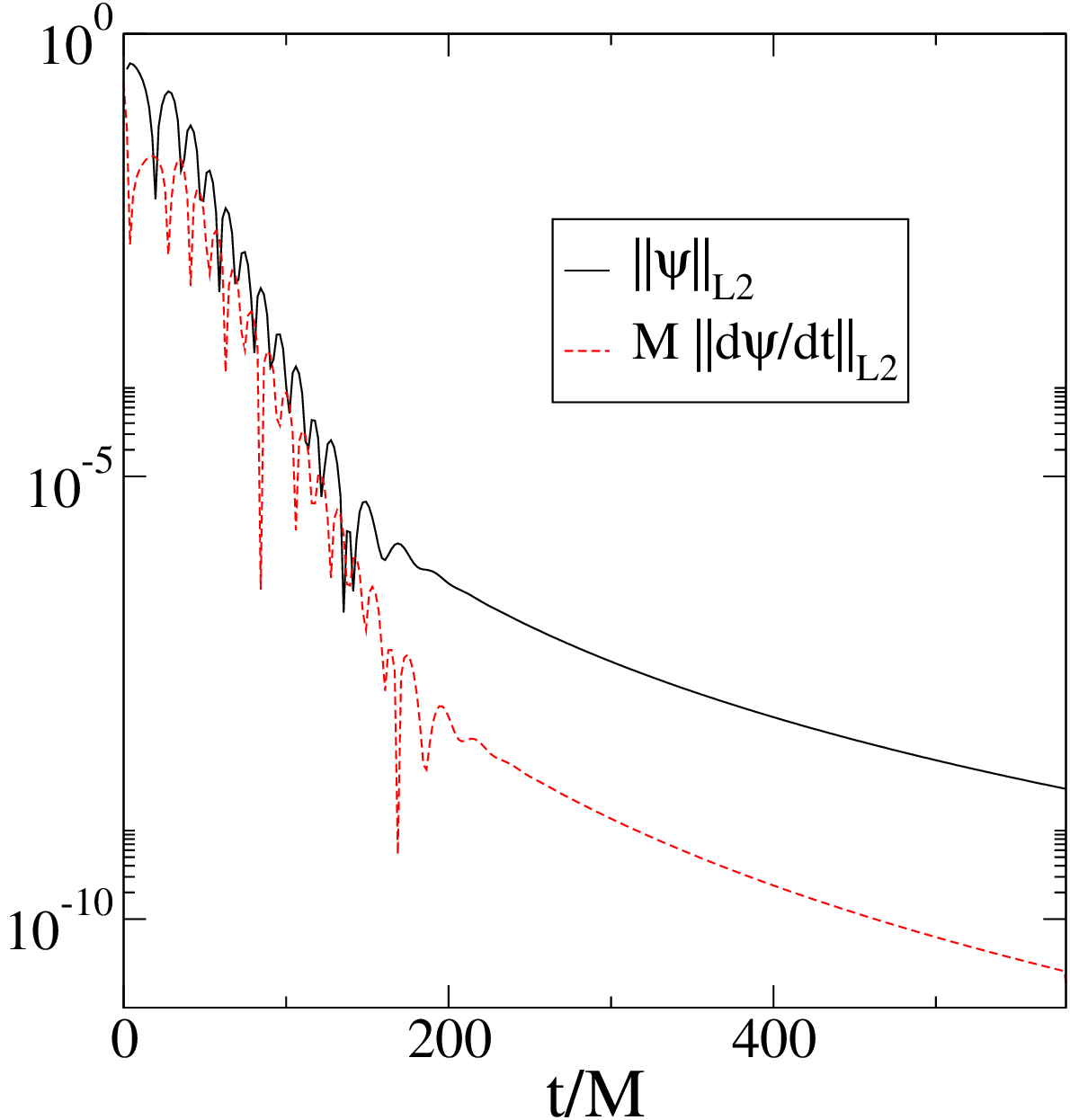}
\end{center}
\caption{The evolution of a scalar field with an initial $Y_{10}$
angular dependence in Schwarzschild spacetime.  Plotted are the
L2 norms of $\psi$ and $\dot\psi$ on a spherical surface of fixed
radius $r$, defined by
$(||f||_{L2})^2 = (1/4\pi)\int f^2 \,d\Omega$. Here the integration
is taken over a surface at $r=11.9M$.  The duration of the initial
burst is about 50$M$.  After the initial burst, the scalar field
settles into the quasinormal ringing phase until about 200$M$ when the
tail phase begins.}
\label{fig:example}
\end{figure}

The evolution of a scalar field in curved spacetime is governed by the
massless Klein-Gordon equation
\begin{equation}
\Box \psi (x,y,z,t) = 0,
\label{wave}
\end{equation}
where $\psi$ is the value of the scalar field and $\Box$ is the
d'Alembertian operator in curved spacetime.  According to
no-hair theorems, the only
nonsingular time-independent solution to Eq.~(\ref{wave}) in a black
hole background is $\psi=0$ everywhere, and
furthermore,
if $\psi$ initially varies in time or space it will evolve until it
reaches this time-independent solution~\cite{Price1972}. 
When observed at a fixed
spatial location as a function of time, the evolution of a scalar
field in a black hole spacetime consists of three distinct phases, as
shown in Fig.~\ref{fig:example}.  The first stage is the initial
burst, which is determined by the initial conditions imposed on the
scalar field. The second stage is the quasinormal ringing phase,
during which outgoing waves interfere with incoming waves that
backscatter off the black hole's potential well.  During this phase
$\psi$ oscillates and decays exponentially, and can be written as a
sum of terms of the form $e^{i\omega_n t}$ for a discrete set of
complex eigenfrequencies $\omega_n$.  During the third stage, or tail
phase, $\psi$ depends on incoming radiation that has been
backscattered off the spacetime curvature at large distances. During
the tail phase, the scalar field 
decays as a power law, $\psi \propto t^{-\mu}$, for the case
of a Schwarzschild background, and there is good
analytical~\cite{BarackOri1999} and numerical~\cite{Krivan1996,Burko2002}
evidence that it decays as a power law for a Kerr background as 
well.

For a scalar field in Schwarzschild spacetime, the power-law decay in
the tail phase is computed using the spherical harmonic decomposition
of the initial data.  The amplitude of each $Y_{\ell m}$ mode present
in the initial data will eventually decay like $t^{-(2\ell+3)}$ at
late times~\cite{Price1972,Gundlach1994,Ching1995,Barack1999a},
assuming that $\psi$ initially falls off quickly enough at 
infinity\footnote{If $\psi$ at large distances is initially a static solution
of Eq.~(\ref{wave}),
then the late-time decay rate~\cite{Price1972} is $t^{-(2\ell+2)}$. 
Note that all static solutions of Eq.~(\ref{wave}) that
are regular at infinity diverge at the 
horizon~\cite{Price1972,Vishveshwara1970}}. 
If one measures
$\psi$ at a single point in space or by some other method that does
not select specific spherical harmonic components, the decay rate
measured at late times will be determined by the smallest $\ell$
present in the initial data, because this is the most slowly decaying
mode.

The late-time behavior of the scalar field becomes more complicated in
Kerr spacetime because of the lack of spherical symmetry.  Although
axisymmetry prevents mixing of spherical harmonics with different $m$
values, harmonics with different values of $\ell$ no longer evolve
independently.  Because of this mode mixing, if the initial data are
proportional to a pure spherical harmonic $Y_{\ell_0 m_0}$, the
evolution should produce spherical harmonics with different values of
$\ell$, and in particular smaller values of $\ell$.  It is not
unreasonable to assume the same power-law time dependence as the
Schwarzschild case, namely $\psi\sim t^{-(2\ell+3)}$, because the
tails are due to radiation backscattered off the weak-field asymptotic
region of spacetime. Given this assumption, the late-time behavior of
the scalar field should be dominated by the smallest value of $\ell$
that is produced by mode mixing, since this is the most slowly
decaying mode.  Given that $\ell\ge m$ and that parity
is conserved (i.e. the equatorial symmetry of the initial data is preserved),
the lowest-order spherical harmonic that may be generated from initial
data proportional to $Y_{\ell_0 m_0}$ is $\ell = m_0$ if $\ell_0-m_0$
is even and $\ell = m_0+1$ if $\ell_0-m_0$ is odd\footnote{The amplitude
of this lowest-order mode may, however, turn out to be zero, in which case
the decay rate would be determined by the lowest-order mode with nonzero
amplitude.}.
  Therefore,
according to this simple picture one would expect at late times
\begin{equation}
\psi \propto \left\{\begin{array}{ll}
             Y_{m_0 m_0}    \,\,t^{-(2m_0+3)},     &\ell_0-m_0 \mbox{ even,}\cr
             Y_{(m_0+1) m_0}\,\,t^{-(2(m_0+1)+3)}, &\ell_0-m_0 \mbox{ odd.}
	         \end{array}
             \right.
\label{eq:naive}
\end{equation}

However, analytical work by Hod~\cite{Hod2000} predicts different behavior
for scalar fields in Kerr spacetime. According to Hod's analysis, the
late-time decay rate does not just depend on the lowest multipole index
$\ell$ permitted by parity and axisymmetry; the initial value
$\ell_0$ also plays a role:  
\begin{equation}
\psi \propto \left\{\begin{array}{ll}
    Y_{\ell_0 m_0}  \,\,t^{-(2\ell_0+3)},    &\ell_0-m_0< 2,    \cr
    Y_{m_0 m_0}     \,\,t^{-(\ell_0+m_0+1)},&\ell_0-m_0\ge 2 \mbox{ (even)},\cr
    Y_{(m_0+1) m_0} \,\,t^{-(\ell_0+m_0+2)},&\ell_0-m_0\ge 2 \mbox{ (odd)}.
	            \end{array}
             \right.
\label{eq:hod}
\end{equation}
This is a deeply surprising result, for it implies that the generated
modes somehow ``remember'' the properties of the initial data that
created them.

It has been recently argued by Poisson~\cite{Poisson2002} that {\it
both\/} Eq.~(\ref{eq:hod}) and the simple picture leading
to Eq.~(\ref{eq:naive})
are valid
descriptions of the late-time dependence of scalar fields in a
rotating spacetime that is weakly curved everywhere; the difference is
merely the choice of spatial coordinates $(r,\theta,\phi)$ used when
setting the initial data.  
Poisson assumes a metric equal to Minkowski space
in spherical coordinates $(r,\theta,\phi)$ plus a stationary
nonspherical perturbation which he treats to linear order.  For
initial data proportional to $f(r) Y_{\ell_0 m_0}(\theta,\phi)$ for some
function $f(r)$, he
finds that the field decays like $t^{-(2 \ell_0 + 3)}$ (there is no
mode mixing to first order in the perturbation).  
He then repeats the calculation using initial data
proportional to $f(r') Y_{\ell_0 m_0}(\theta',\phi')$, where
$(r',\theta',\phi')$ are spheroidal coordinates defined by 
\begin{eqnarray}
\label{eq:spheroidaldef1}
{r^2\sin^2\theta\over{a^2+r'^2}}+{r^2\cos^2\theta\over r'^2} &=& 1,\\
\label{eq:spheroidaldef2}
r'\cos\theta'&=&r\cos\theta,\\
\label{eq:spheroidaldef3}
\phi'&=&\phi,
\end{eqnarray}
for some constant $a$. In this case he finds that the modes
mix because of the nonspherical coordinates,
and the scalar field decays according to Eq.~(\ref{eq:hod}).

Poisson then argues that since radiative falloff is essentially a
weak-field phenomenon, similar conclusions should be true for scalar
fields in a Kerr background, so that coordinate effects would account for
the discrepancy between Eq.~(\ref{eq:naive}) and Eq.~(\ref{eq:hod}).
Consider initial data proportional to $f(r)Y_{\ell_0m_0}(\theta,\phi)$
where $(r,\theta,\phi)$ are any coordinates in which the weak-field
limit of the Kerr metric is spatially isotropic.  Then the only mode
mixing will be due to the strong-gravity region at early times, and at
late times, when the scalar field probes only the weak-gravity region,
each mode that was generated by the mixing will decay like $t^{-(2
\ell + 3)}$. One therefore expects Eq.~(\ref{eq:naive}) to hold.
Now consider initial data proportional to $f(r')Y_{\ell_0
m_0}(\theta',\phi')$ where $(r',\theta',\phi')$ are any coordinates in
which the weak-field limit of the Kerr metric is spheroidal.  Such
coordinates include Boyer-Lindquist coordinates, the coordinates used
in Hod's analysis, which in the weak-field limit reduce to flat space
in spheroidal coordinates $(r',\theta',\phi')$ with the parameter $a$
in Eq.~(\ref{eq:spheroidaldef1}) equal to the Kerr spin parameter.
For such initial data, if strong-gravity mode mixing can be ignored
relative to the mode mixing resulting from the spheroidal coordinate
system (this key assumption is discussed in more detail in
Section~\ref{sec:coordinate-effects}), then the scalar field should
behave according to Eq.~(\ref{eq:hod}).  The ``memory'' effect implied
by Eq.~(\ref{eq:hod}), according to this argument, is due to
coordinates and not physics.

Surprising and seemingly contradictory results have resulted not only
from analytic studies of this problem but also from numerical
simulations.  Early simulations~\cite{Krivan1996} considered cases for
which Eq.~(\ref{eq:naive}) and Eq.~(\ref{eq:hod}) agree, and were
consistent with both predictions, but a more recent
simulation~\cite{Krivan1999} yielded the puzzling result that a scalar
field initially proportional to $Y_{40}$, the lowest multipole mode
for which Eq.~(\ref{eq:naive}) and Eq.~(\ref{eq:hod}) differ, decays
approximately like $t^{-5.5}$, in conflict with both predictions. Most
recently, a 2+1 simulation of an initial $Y_{40}$ mode using ingoing
Kerr coordinates~\cite{Burko2002} agrees with Eq.~(\ref{eq:naive}) to
high accuracy.

Here we solve the scalar wave equation in a fixed Kerr background in
Kerr-Schild coordinates using a 3+1 numerical evolution code.  We
reproduce the known fundamental quasinormal frequency and the known
tail falloff behavior for the Schwarzschild case. We find that for a
black hole with nonzero spin, when we choose initial data proportional
to $f(r) Y_{40}(\theta,\phi)$, where $(r,\theta,\phi)$ are Kerr-Schild
coordinates, we find late-time tail behavior consistent with
Eq.~(\ref{eq:naive}).

We then choose a different set of initial data
proportional to $f(r_{\rm BL}) Y_{40}(\theta_{\rm BL},\phi)$, where
$(r_{\rm BL},\theta_{\rm BL})$ are the Boyer-Lindquist coordinates. Note
that the Boyer-Lindquist coordinates are spheroidal in the sense discussed
earlier and the Kerr-Schild coordinates are spherical, and that
the transformation between $(r,\theta)$ and $(r_{\rm BL},\theta_{\rm
BL})$ is the same
transformation~(\ref{eq:spheroidaldef1}--\ref{eq:spheroidaldef2}) used
by Poisson.  For this initial data, we obtain a quite different
evolution at intermediate times, with different magnitudes of
lower-order spherical harmonics generated during the evolution, even
though the initial data differs from the Kerr-Schild case by a small
amount. However, at very late times it appears that the scalar field
decays according to Eq.~(\ref{eq:naive}).  Our results indicate that
the coordinate effects discussed by Poisson play an important role in
the details of the evolution at intermediate times, but they do not
affect the asymptotic decay rate, presumably because of mode mixing in
the strong-field region, an effect that was not included in Poisson's
analysis.

Because Kerr spacetime is axisymmetric, a 2+1 dimensional
simulation would suffice for the present problem. We work in 3+1
dimensions because we have available a 3+1 dimensional code (see, {\it
e.g.\/}~\cite{Kidder2001,Lindblom2002,Scheel2002}) which is designed
to solve the full nonlinear Einstein evolution equations and is being
used to study the binary black hole problem. This code can be applied
to not only the Einstein equations, but to any first order strongly
hyperbolic system of equations.  For such a numerical code it is
extremely useful to find test problems that are simpler than, for
example, the binary black hole problem, but difficult enough so that
they still provide nontrivial tests of our numerical algorithms.

The simulation of late-time tails is just this type of problem.
Because it is linear and involves fewer dynamical fields, this problem
is simpler than those involving dynamical black holes.  Yet our
treatment of this problem contains many of the features currently
thought to be desirable in a solution of the binary black hole
problem: wave propagation, multiple computational domains,
parallelism, black hole excision with no boundary condition imposed on
the excision surface, and constraint-preserving boundary conditions on
certain fields at the outer boundary. These features will be discussed
further in Section~\ref{sec:numerical-method}.

Although this problem is simpler than the evolution of dynamical black
holes, it is still technically challenging in 3+1 dimensions because
of the requirement for high resolution, long integration times, and a
distant outer boundary.  As discussed in
Section~\ref{sec:numerical-method}, we overcome these difficulties by
the use of multiple computational domains and a pseudospectral
evolution algorithm. The latter yields exponential convergence of
spatial numerical errors for smooth problems, allowing us to achieve a given
level of accuracy using a small fraction of the computational
resources that would be required by a unigrid finite-difference code.

\section{Basic Equations}

\subsection{The Background Spacetime}
\label{sec:background-spacetime}
We write the spacetime metric in the usual 3+1 form
\begin{equation}
ds^2=-\alpha^2\,dt^2+g_{ij}(dx^i+\beta^i\,dt)(dx^j+\beta^j\,dt),
\label{metric}
\end{equation}
where $g_{ij}$ is the three-metric, $\alpha$ is the lapse and
$\beta^i$ is the shift.  The Klein-Gordon equation will involve 
these quantities and also the extrinsic curvature $K_{ij}$,
defined by
\begin{equation}
K_{ij} = -\frac{1}{2\alpha}(\frac{\partial}{\partial t}
-{\cal{L}}_{\vec{\beta}}) g_{ij},
\label{excur}
\end{equation}
where ${\cal{L}}_{\vec{\beta}}$ is a Lie derivative.

The Kerr spacetime is expressed in Kerr-Schild coordinates
$(t,x,y,z)$.  For a Kerr black hole with angular momentum $a M$ in
the $z$ direction, the 3+1 decomposition of the spacetime in
Kerr-Schild coordinates is
\begin{eqnarray}
g_{ij}  &=& \delta_{i j} + 2 H l_i l_j,\label{eq:KSg}\\
\alpha  &=& \left(1+2 H l^t l^t\right)^{-1/2},\label{eq:KSalpha}\\
\beta^i &=& - {2 H l^t l^i \over 1+2H l^t l^t},\label{eq:KSbeta}\\
K_{i j} &=& - \left(1+2 H l^t l^t\right)^{1/2}
	            \left[l_i l_j \d{t} H + 2 H l_{(i} \d{t} l_{j)}\right]
		\nonumber \\
		&&-2\left(1+2 H l^t l^t\right)^{-1/2}
	            \left[\d{(i}\left(l_{j)} H l^t\right)
\right. \nonumber \\ && \left.
+ 2H^2 l^t l^k l_{(i} \d{|k|}l_{j)}
			  + H l^t l_i l_j l^k \d{k} H\right],\label{eq:KSK}
\end{eqnarray}
where $H$ and $l_\mu$ are given in terms of the black hole's mass $M$
and its angular momentum $a M$ by
\begin{eqnarray}
H     &=& {M r_{\rm BL}^3 \over r_{\rm BL}^4 + a^2 z^2},\\
l_\mu &=& \left(1,{r_{\rm BL}x+ay\over r_{\rm BL}^2+a^2},
          {r_{\rm BL}y-ax\over r_{\rm BL}^2+a^2},{z\over r_{\rm BL}}\right),
\end{eqnarray}
and the Boyer-Lindquist coordinate $r_{\rm BL}$ is defined by
\begin{equation}
\label{eq:rdefinition1}
{x^2+y^2\over a^2+r_{\rm BL}^2} + {z^2\over r_{\rm BL}^2} = 1.
\end{equation}
Here and in the following, the quantity $r$ without a BL subscript
refers to the Kerr-Schild radial coordinate defined by
\begin{equation}
r^2 \equiv x^2 +y^2 +z^2.
\end{equation} 

In Kerr-Schild coordinates, the event horizon is located at
\begin{equation} 
r^2 = \left(M+\sqrt{M^2-a^2}\right)^2
 +a^2\left({1-\frac{z^2}{\left(M+\sqrt{M^2-a^2}\right)^2}}\right)^2.
\label{Kerrhorizon}
\end{equation}
Notice that the horizon is not spherical in these coordinates.  We
typically set $a=0.5$; in this case the largest coordinate sphere
contained within the event horizon has a radius of $1.87M$, and the
smallest sphere that is outside the Cauchy horizon has a radius of
$0.52M$.

As a consistency check, we compare results for a Kerr background
with $a=0$ to results using a Schwarzschild background expressed in
Painlev\'{e}-Gullstrand~\cite{painleve21,gullstrand22}
coordinates. In these coordinates, the spatial
three-metric is flat, leading to a simple representation of the
Schwarzschild solution:
\begin{eqnarray}
g_{ij} &=& \delta_{ij},\\
K_{ij} &=& \sqrt{2M/r^3}\delta _{ij} - 3\sqrt{M/2r^3}\hat r_i \hat r_j,\\
\alpha &=&1,\\
\label{eq:paingull2}
\beta^k &=&\sqrt{2M/r}\hat r^k,
\end{eqnarray}
where $\delta _{ij}$ is the Euclidean metric, $r$ is the areal radial
coordinate (which for $a=0$ is the same as the Kerr-Schild and the
Boyer-Lindquist radial coordinate), $\hat r_i=x_i/r$ is the Euclidean unit
vector in the radial direction, and $M$ is the mass of the black
hole. The event horizon is located at $r=2M$.

\subsection{Klein-Gordon equation}

We write the Klein-Gordon equation~(\ref{wave}) in first-order form by
introducing four new variables:
\begin{eqnarray}
\Pi    &\equiv& \frac{-1}{\alpha}\left({\frac{\partial\psi}{\partial t}
	  -\beta^i \frac{\partial\psi}{\partial x^i}}\right), \label{pi}\\
\Phi_i &\equiv& \frac{\partial\psi}{\partial x^i}. \label{phi}
\end{eqnarray}
In the background given by equation~(\ref{metric}), the Klein-Gordon
equation~(\ref{wave}) and the commutivity of partial derivatives yield
the following system of evolution equations
\begin{eqnarray}
\label{evol0}
\frac{\partial \psi}{\partial t} &=& \beta^i \psi_{,i}- \alpha \Pi,\\
\frac{\partial \Pi}{\partial t}  &=& \beta^i \Pi_{,i} -
\alpha g^{ij}\Phi_{i,j} + \alpha g^{ij}\Gamma^k_{ij}\Phi_k 
\nonumber \\ 
&& -g^{ij}\Phi_j\alpha_{,i} +\alpha K\Pi,\label{evol1} \\
\frac{\partial \Phi_i}{\partial t} &=& \beta^j \Phi_{i,j}+\Phi_j
                      \beta^j{}_{,i} - \alpha\Pi_{,i} - \Pi \alpha_{,i}.
\label{evol2}
\end{eqnarray}
where $A_{,b}$ indicates differentiation with respect to $x^b$.

The system~(\ref{evol0}--\ref{evol2}) is symmetric hyperbolic,
so the quantities $\psi$, $\Pi$ and
$\Phi_i$ may be decomposed into characteristic fields that propagate
with well-determined characteristic speeds with respect to any
two-dimensional surface, such as a boundary.  If the normal to the
surface is $\xi_i$, then the characteristic fields are
\begin{eqnarray}
 u^{\pm} &=& \Pi \pm \xi^i\Phi_i,\\
 u^{0}_i &=& \Phi_i - \xi_i\xi^j \Phi_j,\\
 u^{\psi}&=& \psi.
\end{eqnarray}
The fields $u^{\pm}$ propagate along null rays (coordinate velocity
$v^i=-\beta^i \pm \alpha \xi^i$), and the other fields propagate along
the normal to the spatial hypersurface (coordinate velocity
$v^i=-\beta^i$).  Note that all characteristic fields propagate
causally.  The decomposition into characteristic fields is invaluable
for the purpose of setting mathematically consistent boundary
conditions.  At a boundary with normal $\xi_i$, boundary conditions
must be imposed only on incoming characteristic fields, that is, those
having $v^i\xi_i<0$. Boundary conditions must not be imposed on other
characteristic fields.

Note that the definition of $\Phi_i$, Eq.~(\ref{phi}), becomes a set
of constraints,
\begin{equation}
C_i = \frac{\partial\psi}{\partial x^i} - \Phi_i,
\label{constraint}
\end{equation}
that must be satisfied at all times.  If $C_i=0$ initially and the
solution is advanced in time by solving
Eqs.~(\ref{evol0}--\ref{evol2}) exactly, then $C_i$ will remain zero
for all times, as long as the boundary conditions are
consistent with $C_i=0$. However, both numerical truncation errors 
and boundary errors can cause $C_i$
to drift away from zero. Therefore, tracking the evolution of $C_i$
provides a test of the accuracy of our simulations.

\section{Numerical Method}
\label{sec:numerical-method}
\subsection{Computational Domain}
\label{sec:computational-domain}
We solve Eqs.~(\ref{evol0}--\ref{evol2}) in a 3D spherical shell
extending from a radius $r=r_{\rm min}$ lying between the event
horizon and the Cauchy horizon to some large radius $r=r_{\rm max}$.
Because all characteristic fields propagate causally, placing the
inner boundary inside the event horizon means that all characteristic
fields are outgoing (into the hole) there: $v^i\xi_i>0$. Therefore we
impose no boundary condition at the inner boundary.  Typically we
choose $r_{\min}=1.75M$.

The outer boundary must be placed at a large radius because the
power-law tails of interest are due to backscattering of radiation off
the background geometry at large distances.  If we wish to measure the
tail contribution to the scalar field at time $t$ and radius $r$, then
the outer boundary must be placed roughly at $r_{\rm max} > r + t/2$,
so that the backscattering responsible for the tail contribution
occurs within our computational domain.  Because determining the decay
rate of tails requires evolution to approximately $t=600M$, we
typically place our outer boundary at $r_{\rm max}=300M$.

To facilitate multiprocessing, the domain is divided into concentric
subdomains, each a spherical shell with a width of $10M$.  Each
subdomain is evolved independently except for boundary conditions, so
each subdomain can be evolved on a different processor, with
interprocessor communication occurring only at the boundaries.  To
impose boundary conditions at an interdomain boundary, we set the time
derivative of each incoming characteristic field equal to the time
derivative of the corresponding outgoing field from the neighboring
subdomain.

\subsection{Solution Technique}

Our numerical methods are essentially the same as those we have
applied to the evolution problem in full general
relativity~\cite{Kidder2000a,Kidder2001,Lindblom2002,Scheel2002}.
We use a pseudospectral technique on each subdomain to evolve
Eqs.~(\ref{evol0}--\ref{evol2}) in time.  Given a system of partial
differential equations
\begin{equation} 
\frac{\partial}{\partial t} f(\mathbf{x},t) = 
{\cal F}(f(\mathbf{x},t),\partial f(\mathbf{x},t)/\partial x^i),
\label{diffeq}
\end{equation}
where $f$ is a vector of variables, the solution $f(\mathbf{x},t)$ is
expressed as a time-dependent linear combination of $N$ basis
functions $\phi(\mathbf{x})$:
\begin{equation}
f_N(\mathbf{x},t) = 
	\sum_{k=0}^{N-1}\tilde{f}_k(t) \phi_k(\mathbf{x}).
\label{decom}
\end{equation}
Spatial derivatives are evaluated analytically using the known
derivatives of the basis functions:
\begin{equation}
\frac{\partial}{\partial x^i} f_N(\mathbf{x},t) 
= \sum_{k=0}^{N-1}\tilde{f}_k(t)
  \frac{\partial}{\partial x^i}\phi_k(\mathbf{x}).
\label{decomderiv}
\end{equation}

The coefficients $\tilde{f}_k(t)$ are chosen so that equation
(\ref{diffeq}) is satisfied exactly at $N_c$ collocation points
selected from the spatial domain.  The values of the coefficients are
obtained by the inverse transform
\begin{equation}
\tilde{f}_k(t) = \sum_{i=0}^{N_c-1}f_N(\mathbf{x}_i,t)
                       \phi_k(\mathbf{x}_i) w_i, 
\label{invdecom}
\end{equation}
where $w_i$ are weights specific to the choice of basis functions and
collocation points. One can now transform at will, using equations
(\ref{decom}) and (\ref{invdecom}), between the spectral coefficients
$\tilde{f}_k(t)$ and the function values at the collocation
points $f_N(\mathbf{x}_i,t)$.  The differential equations
(\ref{diffeq}) are now rewritten, using equations
(\ref{decom}--\ref{invdecom}), as a set of {\it
ordinary\/} differential equations for the function values at the
collocation points,
\begin{equation} 
\frac{\partial}{\partial t} f_N(\mathbf{x}_i,t) 
                      = {\cal G}_i (\{f_N(\mathbf{x}_j,t)\}),
\label{odiffeq}
\end{equation}
where ${\cal G}_i$ depends on $f_N(\mathbf{x}_j,t)$ for
all $j$.

Equations~(\ref{odiffeq}) are integrated in time using a fourth order
Runge-Kutta  method.  Boundary  conditions are incorporated into the
right-hand side of Eqs.~(\ref{odiffeq}) using the technique of
Bj{\o}rhus~\cite{Bjorhus1995}: if $P^{+}$ is the projection operator that
annihilates all incoming characteristic fields at a boundary, and $P^{-}$ is
$1-P^{+}$, then at
each boundary point $i$ the differential equation~(\ref{odiffeq})
is modified as follows:
\begin{equation} 
\frac{\partial}{\partial t} f_N(\mathbf{x}_i,t) 
                          = P^+ {\cal G}_i (\{f_N(\mathbf{x}_j,t)\})
			   +P^- {\cal B}_i(\{f_N(\mathbf{x}_j,t)\}),
\label{odiffeqbdry}
\end{equation}
where $P^-{\cal B}_i(\{f_N(\mathbf{x}_j,t)\})$ encodes the boundary
condition placed on the time derivatives of the
incoming characteristic fields.

For computational subdomains with spherical boundaries, it is natural to
use spherical coordinates. We choose our basis functions to
be Chebyshev polynomials in radius and spherical
harmonics in angles.
Although our basis functions are based on spherical coordinates, we
choose our dynamical scalar field variables and our gravitational
variables to be the {\it Cartesian \/} components, and not the
spherical components, of the relevant quantities.  This allows us to
use the same angular basis functions for all variables without regard
to regularity.

To eliminate high-frequency numerical instabilities that sometimes
develop during our simulations, we apply a filter to the right-hand
sides of Eqs.~(\ref{odiffeq}) before incorporating boundary conditions
via the Bj{\o}rhus algorithm. The filter consists of simply setting
high-frequency spherical harmonic coefficients to zero.  The
components that are set to zero depend on which equation is being
solved: if $\ell_{\rm max}$ is the index of the highest frequency
basis function $Y_{\ell m}$, then typically the largest $\ell$
retained in the right-hand side of the $\Pi$ equation~(\ref{evol1}) is
$3\ell_{\rm max}/2-1$, and the largest $\ell$ retained in the
right-hand sides of the $\Phi_i$ equations~(\ref{evol2}) is
$3\ell_{\rm max}/2$.  This is similar to the ``$3/2$ rule'' commonly
used to eliminate nonlinear aliasing errors~\cite{Canuto1988}. No
filtering is done for the $\psi$ evolution equation~(\ref{evol0}), and
filtering is not performed on the radial basis functions.  The degree
of filtering necessary to obtain stability depends on
both the background geometry and the configuration of subdomains,
and is not completely understood. For example, in some cases no filtering is
needed, and in others it suffices to zero only modes with
$\ell=\ell_{\rm max}$ in the $\Phi_i$ equations~(\ref{evol2}) and
modes with $\ell\ge\ell_{\rm max}-1$ in the $\Pi$
equation~(\ref{evol1}).

\subsection{Outer boundary conditions}

The simplest outer boundary condition is obtained by setting the time
derivatives of the incoming characteristic fields $u^{-}$, $u^{0}_i$,
and $u^{\psi}$ to zero.  While this works well for a Schwarzschild
background in Painlev\'{e}-Gullstrand coordinates, for a Kerr
background in Kerr-Schild coordinates, even for $a=0$, this boundary
condition produces strong violations of the constraint $C_i$ at the
outer boundary, even at $t=0$. These constraint violations propagate
inward and grow, eventually dominating the numerical solution. Because these
constraint violations appear as oscillations in the variable $\psi$
but do not affect the fields $\Pi$ and $\Phi_i$, we were able to
greatly reduce them by changing the boundary condition on
$u^\psi$ to
\begin{equation}
\frac{\partial u^{\psi}}{\partial t} = -\alpha\Pi + \beta^i \Phi_i.
\label{boundary}
\end{equation}
This is the same as Eq.~(\ref{evol0}) except that $\psi_{,i}$ has been
replaced by $\Phi_i$ using the constraint~(\ref{constraint}). Exactly
this type of boundary condition has been used
before~\cite{Calabrese2001} in the field of numerical relativity, where
finding methods of constructing boundary conditions that preserve the
constraints is a topic of active 
investigation~\cite{FriedrichNagy1999,
                    Calabrese2002c,Szilagyi2002,Frittelli2003a,Frittelli2003b}.

\begin{figure} 
\begin{center}
\includegraphics[width=3.0in]{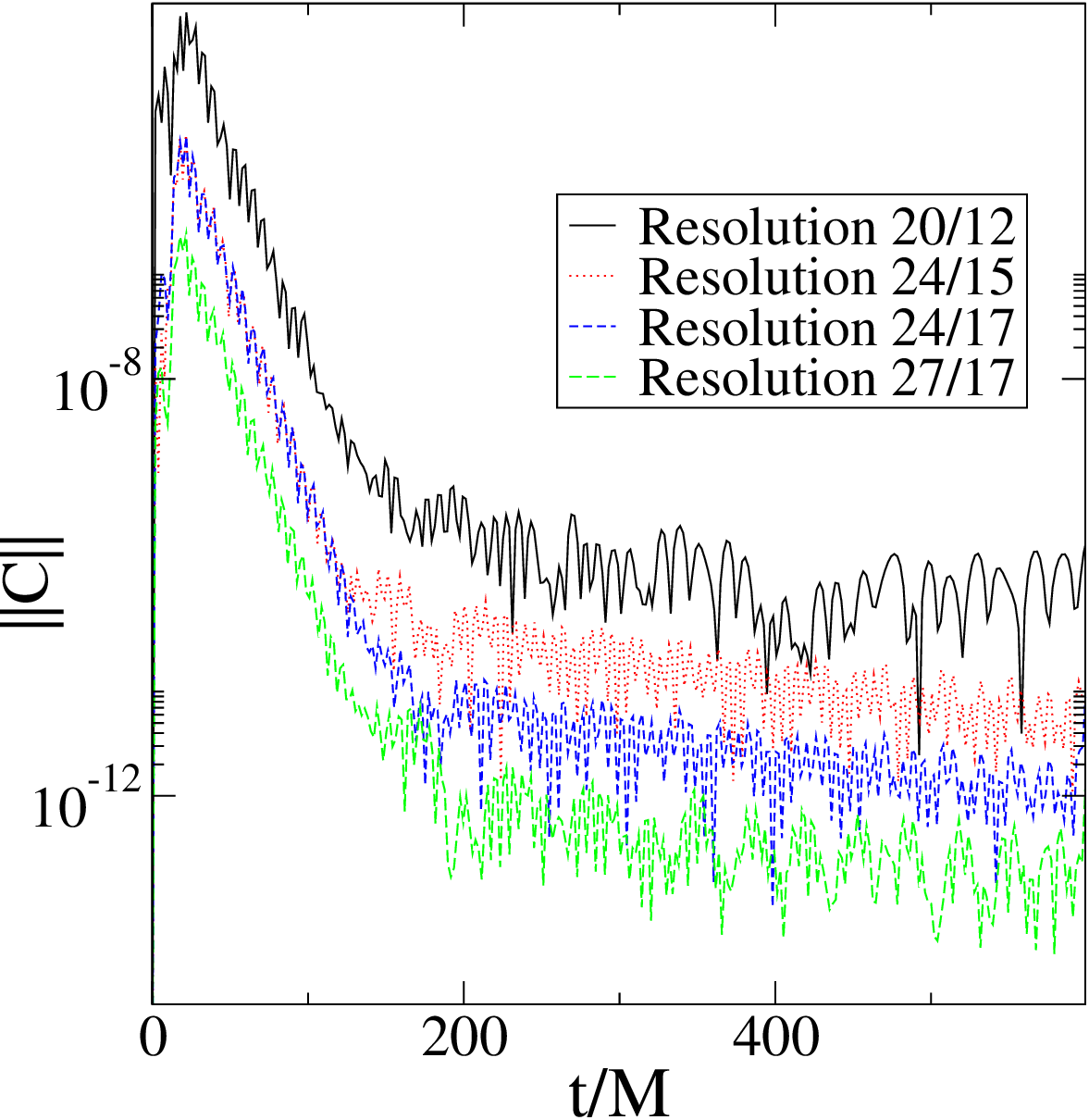}
\end{center}
\caption{Constraint violation during a scalar field evolution for
four radial/angular resolutions (the notation $R/\ell_{\rm max}$ means that we
use $R$ radial basis functions and retain angular basis functions up
to order $\ell_{\rm max}$).  Plotted is $||C||$, where
$(||C||)^2 = (1/4\pi)\int C_i C^i \,d\Omega$. The integration
is taken over the surface $r=11.75M$.
For comparison, the value of the scalar field at $t=300M$ at
$r=11.75M$ is on the order of $10^{-8}$. This evolution corresponds to
the $Y_{20}$ case shown in Figures~\ref{KerrRing} and~\ref{KerrPow}.}
\label{constraintfig}
\end{figure}

Figure~\ref{constraintfig} shows the norm of the constraint for
several different resolutions during an evolution of a scalar field in
Kerr spacetime, using the boundary condition~(\ref{boundary}).  The
constraint violations decrease rapidly with increasing resolution.
For the higher resolutions, the constraint violation is small compared
to the magnitude of the scalar field.

Even with the use of Eq.~(\ref{boundary}), reflections (with
a small constraint-violating contribution) occur when an outgoing pulse of
scalar field reaches the outer boundary.  The reflected pulse then
propagates (causally) inward. 
These reflections can be reduced by modifying the
boundary condition on $u^{-}$:
\begin{equation}
\label{boundary2}
\frac{\partial u^{-}}{\partial t} = -\Pi/r.
\end{equation}
This is equivalent to assuming the Sommerfeld condition $\psi =
f(t-r)/r$, for some unknown function $f$, at the outer boundary. In
practice, imposing this boundary condition proved less critical than
imposing Eq.~(\ref{boundary}).  This is because, as explained in
Sec.~\ref{sec:computational-domain}, for studying tails our
integration time is less than the time it takes light to travel from
the black hole to the outer boundary and back again.  Therefore,
although it is important that the outer boundary condition is
well-behaved when there is no wave there, the magnitude and nature of
the reflections produced when a wave passes through the boundary is
largely irrelevant, because the evolution ends before these
reflections reach the observation point.

\subsection{Convergence}

The convergence properties of a pseudospectral code is more difficult
to analyze than, for instance, a second-order finite-difference
code. This is because in the former there are several sources of
truncation error that scale differently with resolution. Spatial
truncation errors should converge exponentially (and errors associated
with the radial direction may have a different exponential convergence
rate than those associated with the angular directions because
different basis functions are used). Time integration errors should
scale like $(\Delta t)^{4}$ because we are using a fourth-order
Runge-Kutta time integrator.  Furthermore, the Courant condition
constrains $\Delta t$ as a function of spatial resolution. For the
resolutions we use, the scaling is roughly ${\rm max}(\Delta t) \sim
N_r^{-2}$, where $N_r$ is the number of radial collocation points; the
scaling is not simply $\Delta t\sim N_r^{-1}$ because the collocation
points are distributed nonuniformly.

\begin{figure} 
\begin{center}
\includegraphics[width=3.0in]{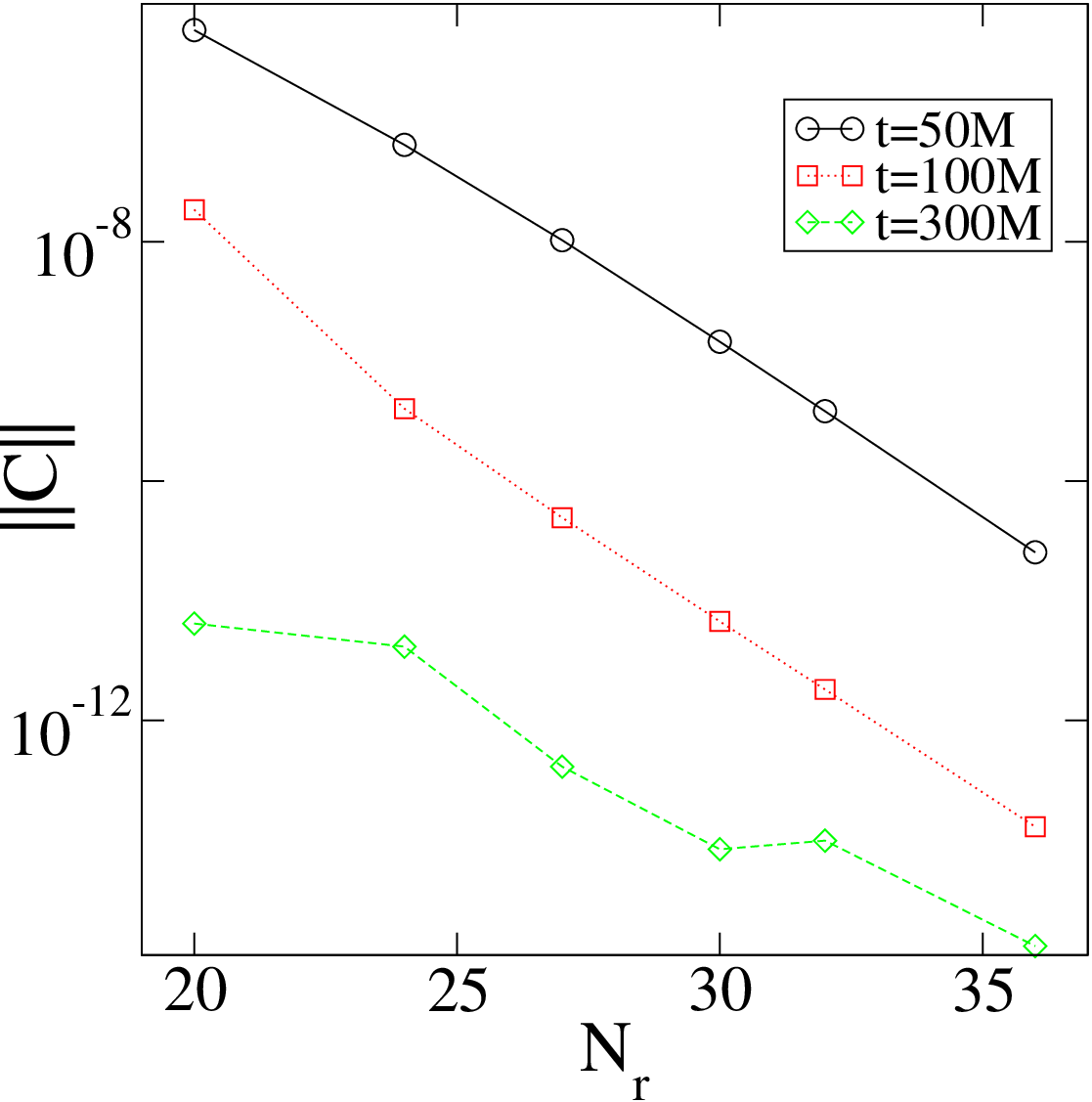}
\end{center}
\caption{The same constraint norm as shown in Figure~\ref{constraintfig},
but plotted versus the number of radial collocation points for
three different values of $t$.
The angular resolution is fixed at $\ell_{\rm max}=17$.
These evolutions correspond to
the $Y_{20}$ case shown in Figures~\ref{KerrRing} and~\ref{KerrPow}.}
\label{fig:radialconvergence}
\end{figure}

\begin{figure} 
\begin{center}
\includegraphics[width=3.0in]{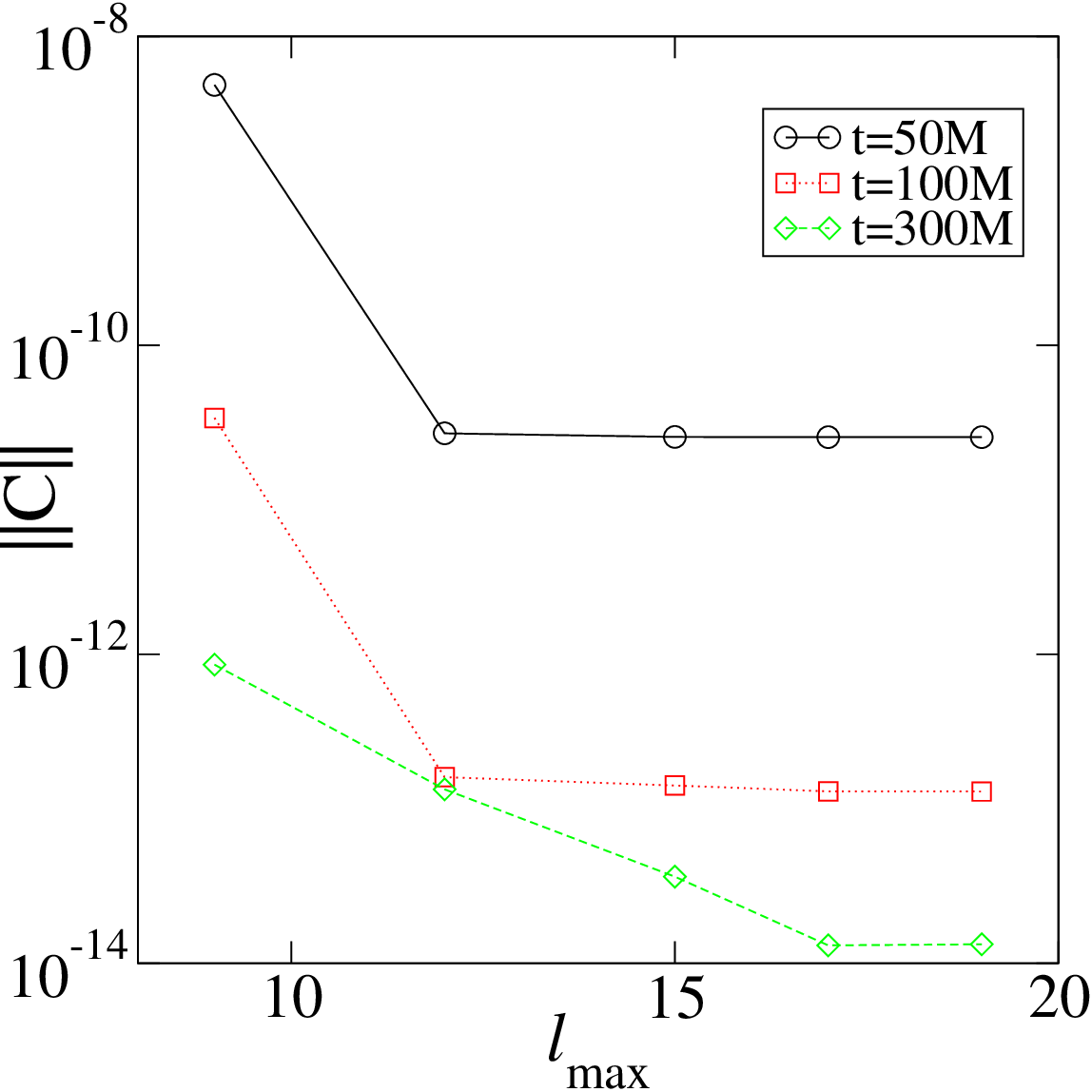}
\end{center}
\caption{The same constraint norm as shown in Figure~\ref{constraintfig},
but plotted versus the angular resolution $\ell_{\rm max}$ for
three different values of $t$. The radial resolution is fixed at $N_r=36$
and $\Delta t$ is fixed at $0.055M$.
These evolutions correspond to
the $Y_{20}$ case shown in Figures~\ref{KerrRing} and~\ref{KerrPow}.}
\label{fig:angularconvergence}
\end{figure}

Figure~\ref{fig:radialconvergence} shows the norm of the constraint as
a function of radial resolution at different times, for the evolutions
shown in Figure~\ref{constraintfig}. The angular resolution is fixed but
$\Delta t$ is varied so that the Courant condition remains
satisfied. The convergence is exponential, indicating that the radial
spatial errors dominate both the time integration errors and the
angular integration errors. Even at late times, when the scalar field
is very small, the convergence plot is still roughly exponential,
although it is noisier than at early times.
Figure~\ref{fig:angularconvergence} shows the norm of the constraint
as a function of angular resolution at different times, for fixed
(high) radial resolution and fixed $\Delta t$.  At late times, the
convergence is exponential for low resolution and then saturates when
the angular truncation error drops beneath radial truncation error.
For early times the angular truncation error is already small, even at
low resolution (as is expected for initial data that is pure $\ell=2$
without higher angular components), so except for the difference in
the two lowest resolutions one does not see any dependence on the
number of angular collocation points.  Presumably, for large enough
spatial resolution, the fourth-order time integration error should
dominate (unless the errors drop beneath roundoff level first), but we
do not see this for the resolutions considered here.

\section{Results and Discussion}
\label{sec:results-discussion}
\subsection{Schwarzschild}
\label{sec:schwarzschild}
As a test of our numerical techniques, we began by evolving the
well-understood case of a scalar field in a Schwarzschild background.
We write the background in Painlev\'e-Gullstrand coordinates, and we
choose initial data of the form
\begin{eqnarray}
\psi   &=& 0,\\
\Phi_i &=& 0,\\
\Pi    &=& \Pi_0(r,\theta,\phi)\equiv
	   e^{-(r-r_0)^2/w^2}Y_{\ell_0 m_0}(\theta,\phi).
\label{initial}
\end{eqnarray}

Figure~\ref{fig:example} shows the results of a simulation with
$\ell_0=1$, $m_0=0$. For all simulations shown here we set $r_0=12M$
and $w = 2M$.  Plotted are results obtained using resolution 24/9
(where the notation $R/\ell_{\rm max}$ means that we use $R$ radial basis
functions and retain angular basis functions up to order $\ell_{\rm max}$).
From roughly $t=40M$ to $140M$ the scalar field behaves like $\psi
\sim e^{-i\omega t}$ with $\omega M \sim 0.29 - 0.097 i$.  This agrees
with published values~\cite{Froman1992} of the least-damped
quasinormal frequency for scalar $\ell=1$ perturbations of the
Schwarzschild geometry.

As time increases, the decay of the scalar tails approaches the
expected power-law decay $\psi \propto t^{-\mu}$.  Since we cannot
numerically evolve the scalar field out to infinite time, our results
do not exactly match the analytically predicted power law.  To
facilitate the determination of the power law governing the decay of
the scalar field during the tail phase, the scalar field and its time
derivative were combined into a ``numerical power index'',
following~\cite{burko_ori97}:
\begin{equation}
\mu_N \equiv \frac{-t ||\dot\psi||_{L2}}{||\psi||_{L2}},
\label{pindex}
\end{equation}
where the L2 norms are defined by
\begin{equation}
(||f||_{L2})^2 = (1/4\pi)\int f^2 \,d\Omega,
\end{equation}
with integration over a surface of fixed $r$.
We compute $\dot\psi$ using Eq.~(\ref{pi}) rather than
taking numerical time derivatives of $\psi$.

At finite times during the tail phase, the scalar field behaves like
\begin{equation}
\psi \propto t^{-\mu} + {\cal{O}}(t^{-(\mu+1)}) + \ldots,
\end{equation}
which implies that
\begin{equation}
{\mu}_N = \mu + {\cal{O}} (t^{-1}) + {\cal{O}} (t^{-2}) + \ldots.
\label{pindevol}
\end{equation}
At late times, the power index asymptotically approaches $\mu$.

\begin{figure} 
\begin{center}
\mbox{\includegraphics[width=3.0in]{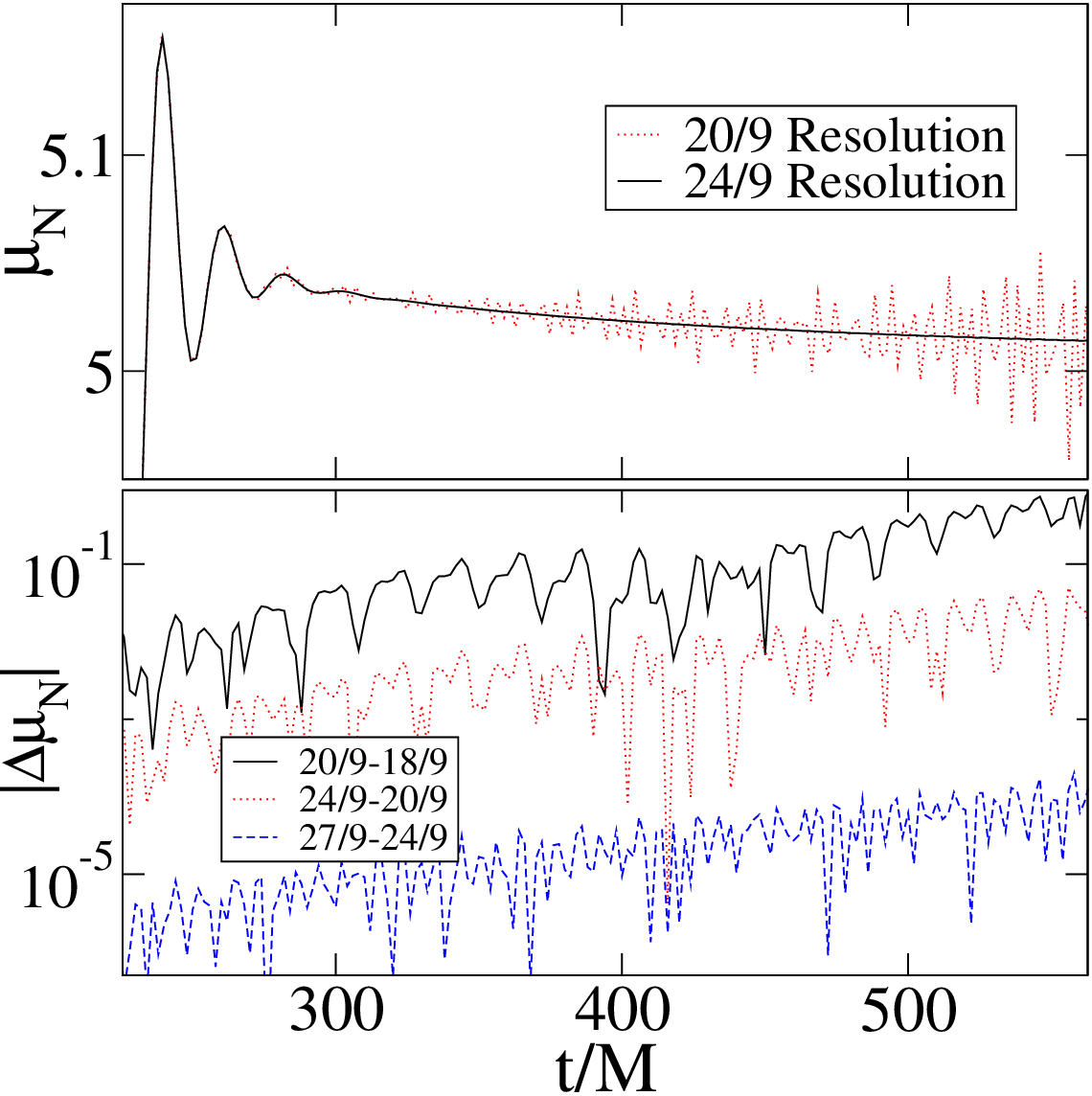}}
\end{center}
\caption{The top graph shows the evolution of the power index $\mu_N$
evaluated at $r=11.9M$ as a function of time for the case shown in
Figure~\ref{fig:example}, and for the same case run at a lower
resolution. The bottom graph shows differences in $\mu_N$
between runs done at different resolutions.}
\label{PGY10}
\end{figure}

Figure~\ref{PGY10} shows the evolution of the power index for two
different radial resolutions. The power index is approaching a value
of five, which corresponds to the predicted $t^{-(2l+3)}$ decay rate.
Moreover, as the resolution increases, the numerical results converge,
as can be seen from the bottom graph in Figure~\ref{PGY10}.

According to Eq.~(\ref{pindevol}), we can obtain a better estimate for
$\mu$ by performing a least-squares polynomial fit to $\mu_N$ as a
function of $t^{-1}$.  The least-squares fit will also give us an
error estimate~\cite{numrec_c} for $\mu$ as long as we provide error
estimates for each of our numerical values of $\mu_N$.  Our code
provides $\mu_N$ at a discrete set of time values $t_i$.  To estimate
the error in $\mu_N$ at time $t_i$ we use
\begin{equation}
\delta \mu_N(t_i) = \max_{j\in \{|i-j|\le W\!\}} \left\{
	\left|\mu_N(t_j;{\rm hi})-\mu_N(t_j;{\rm lo})\right|\right\},
\label{eq:errorestimate}
\end{equation}
where $\mu_N(t_j;{\rm hi})$ and $\mu_N(t_j;{\rm lo})$ represent
numerical values for $\mu_N$ at the highest and next-highest spectral
resolution that we used. The purpose of maximizing the error over neighboring
points is
to treat the cases in which $\mu_N(t_j;{\rm hi})$ and
$\mu_N(t_j;{\rm lo})$ spuriously agree at a single point---without the
maximization this point would have an artificially small error
estimate. The size $W$ of the maximization window is typically $I/20$,
where $I$ is the total number of discrete values of $\mu_N(t_i)$ that
we use for the fit.  Near the points $t_0$ and $t_I$ we translate the
maximization window in Eq.~(\ref{eq:errorestimate}), so that, for
example, for $i=0$ the window goes from $j=0$ to $j=2W+1$.

Using a linear fit to the form~(\ref{pindevol}) we obtain $\mu=4.99\pm
0.01$, and using a quadratic fit we obtain $\mu=5.00\pm 0.08$. These
agree with the accepted value to within about a percent.  We perform
the fits only for data in the tail region of Figure~\ref{PGY10}, that
is, for $t>400M$.  The estimate of $\mu$ is relatively insensitive to
the exact region of $t$ in Figure~\ref{PGY10} that we choose to
perform the fit.

\subsection{Kerr}
\label{sec:kerr-background}

\begin{figure} 
\begin{center}
\mbox{\includegraphics[width=3.0in]{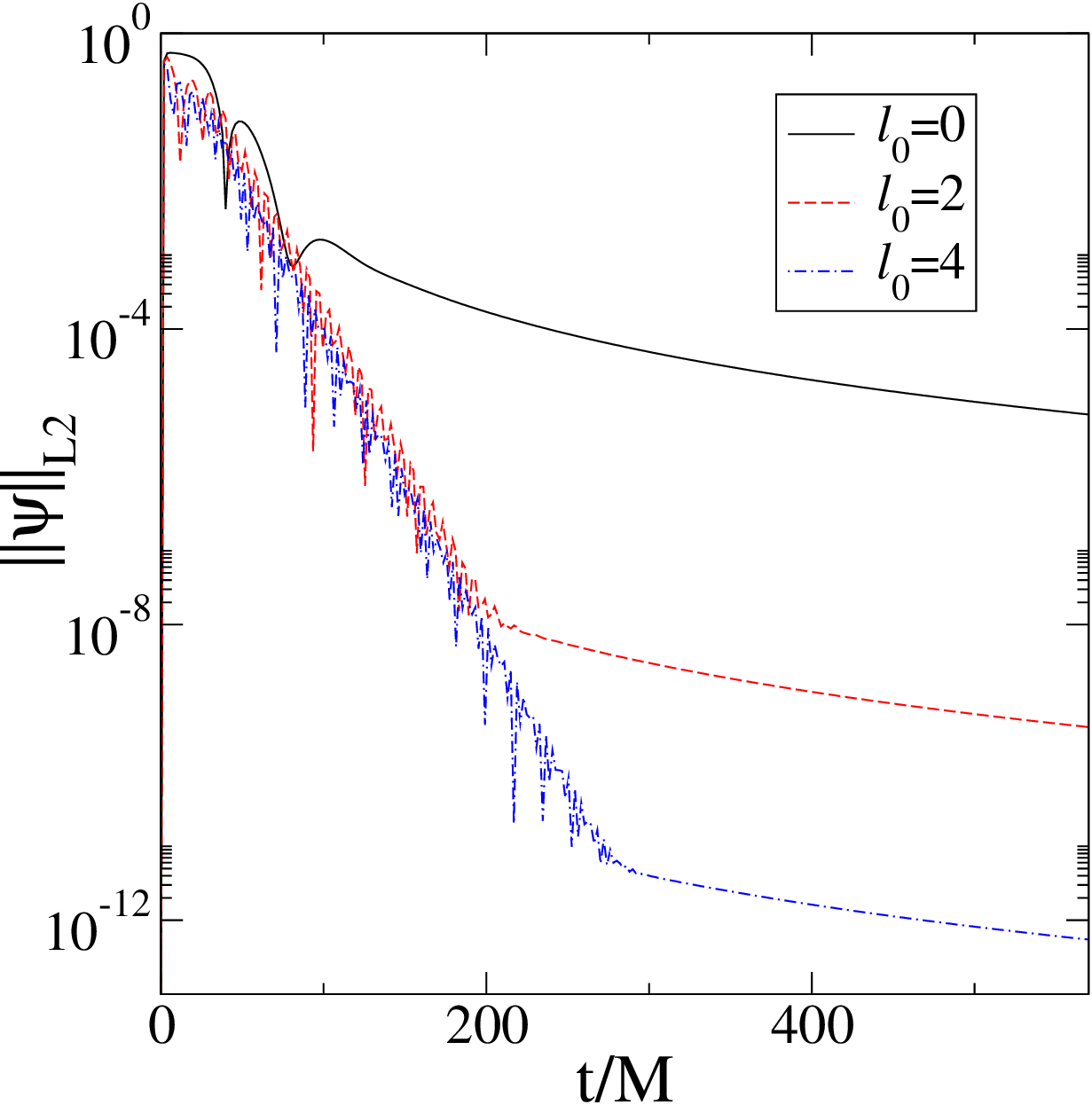}}
\end{center}
\caption{Evolution of a scalar field for three different cases:
initial data proportional to $Y_{00}$, $Y_{20}$, and $Y_{40}$. Plotted
are the L2 norms of $\psi$ evaluated on the surface $r=11.75M$. The
resolution is 20/12 for $\ell_0=0$, 24/17 for $\ell_0=2$, and 27/26
for $\ell_0=4$. Higher resolution is needed for larger $\ell_0$ in
order to resolve the much smaller late-time tail.}
\label{KerrRing}
\end{figure} 

\begin{figure} 
\begin{center}
\mbox{\includegraphics[width=3.0in]{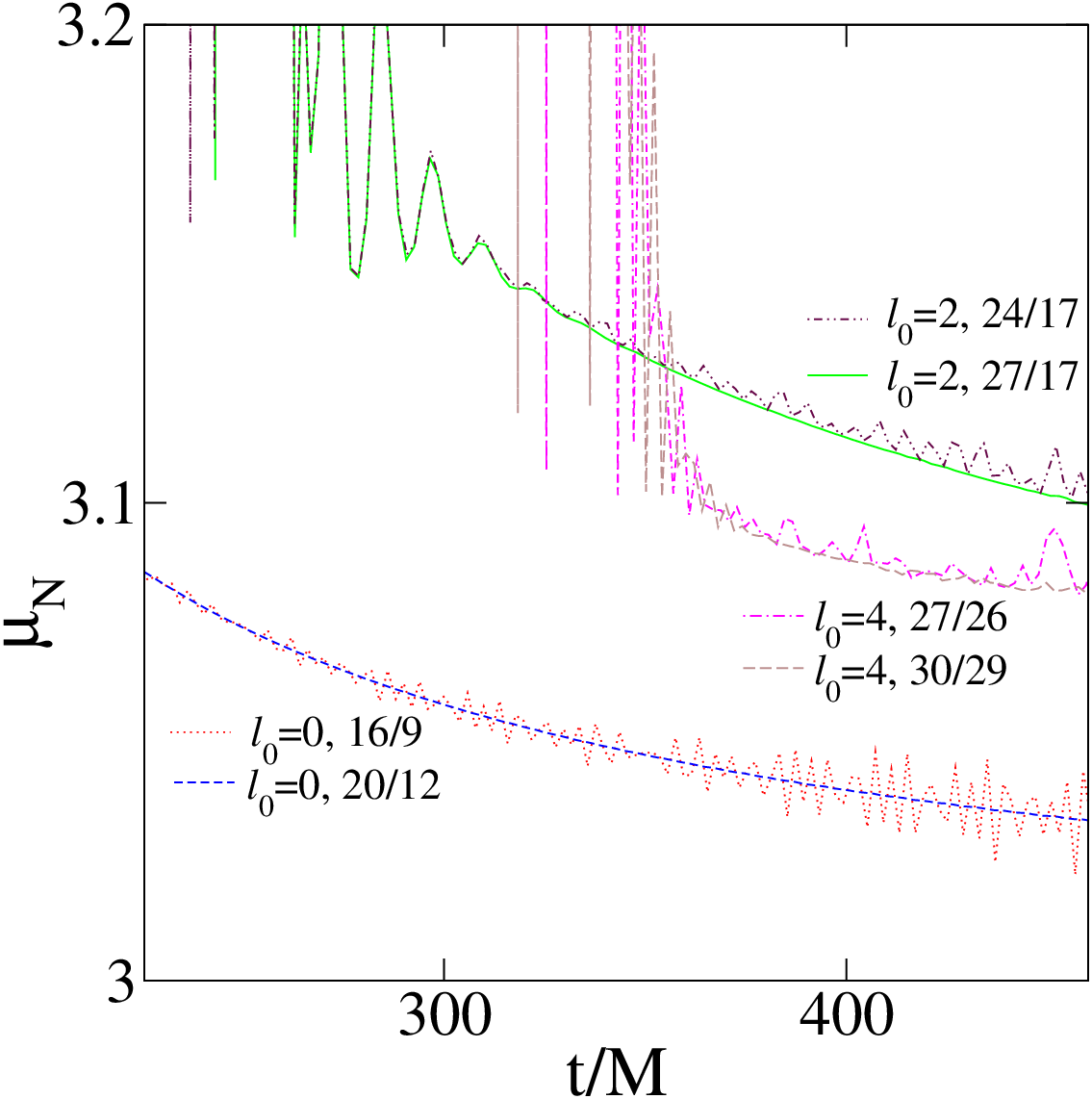}}
\end{center}
\caption{The evolution of the power index $\mu_N$ evaluated at
$r=11.75M$ for the same cases as Fig.~\ref{KerrRing}.  The power
indices appear to asymptotically approach a value of three.  For each case,
two resolutions are shown to demonstrate convergence.}
\label{KerrPow}
\end{figure}

Following our numerical trials with a Schwarzschild background, we
turned our attention to scalar fields around rotating black holes.
For our background spacetime we used a Kerr geometry with spin
$a=0.5M$.  Figure~\ref{KerrRing} displays the evolution of the scalar
field on this Kerr background for three different choices of initial data.
The initial data are taken to have the form~(\ref{initial}), where
now $(r,\theta,\phi)$ are related to the Kerr-Schild coordinates
$(x,y,z)$ defined in Section~\ref{sec:background-spacetime} in the
usual way
\begin{eqnarray}
x &=& r \sin\theta\cos\phi,\\
y &=& r \sin\theta\sin\phi,\\
z &=& r \cos\theta.
\end{eqnarray}
The top, middle, and bottom plots in the figure show the decay of a
scalar field initially proportional to $Y_{00}$, $Y_{20}$, and
$Y_{40}$, respectively.  Since the latter two cases have the same
initial value of $m_0=0$ and are of even parity, we expect that a
$Y_{00}$ mode will be generated during these evolutions. Thus,
according to the simple mode-mixing picture discussed above, all three
evolutions in Figure~\ref{KerrRing} should follow the same power-law decay
at late times.  Figure~\ref{KerrRing} supports this prediction:
although the quasinormal ringing phases are dissimilar, the slope of
the tails do appear to match.  

The evolution of the power index for these cases is shown in
Figure~\ref{KerrPow}.  The power indices approach a value of three,
which corresponds to the Price decay rate formula $\psi\sim
t^{-(2\ell+3)}$ for an $\ell=0$ mode.  Estimates of $\mu$ obtained by
least-squares fits to the numerical data can be found in
Table~\ref{table}.  These estimates all fall within a few percent of
the value $\mu=3$.

The tails have a much smaller magnitude for evolutions with larger
initial values $\ell_0$. For example, for $\ell_0=4$, the scalar field
approaches a magnitude of $10^{-12}$ at late times, and its time
derivative is two orders of magnitude smaller. Thus we are forced to
use larger number of spectral coefficients to resolve the large
$\ell_0$ cases. However, because the accuracy of pseudospectral
methods increases exponentially with the number of collocation points,
increasing the number of coefficients only by roughly a factor of two
in each dimension was sufficient to resolve even the $\ell_0=4$ case.

Note that the $\ell_0=4$ case shown in Figures~\ref{KerrRing}
and~\ref{KerrPow} is the smallest value of $\ell_0$ for which the two
analytical predictions, Eqs.~(\ref{eq:naive}) and~(\ref{eq:hod}),
disagree.  Our results support the simple picture leading
to~(\ref{eq:naive}), which yields a $t^{-3}$ falloff for this case,
rather than Eq.~(\ref{eq:hod}), which predicts a $t^{-5}$ falloff.

\begin{table}
\begin{tabular}{ccrcl@{$\pm$}l}
$\ell_0$&$m_0$&$(r,\theta)$&$a/M$&\multicolumn{2}{c}{$\mu$}\\
\hline
1&0&PG&0&5.00    &0.08  \\
0&0&KS&0.5&2.989 &0.005 \\
2&0&KS&0.5&3.00  &0.006 \\
2&1&KS&0&6.99    &0.03  \\
2&1&KS&0.5&6.99  &0.04  \\
3&1&KS&0.5&5.23  &0.19  \\
4&0&KS&0.5&3.001 &0.003 \\
4&0&BL&0.5&2.8   &0.3   \\
\end{tabular}
\caption{Numerically-determined power-law decay rates. Shown are the
spherical harmonic indices of the initial data, the $(r,\theta)$
coordinates used in the initial data function $\Pi_0$ in Eq.~(\ref{initial})
(Painlev\'{e}-Gullstrand, Kerr-Schild, or
Boyer-Lindquist), the spin of the hole,
and the best-fit power index at late times.}
\label{table}
\end{table}

Table~\ref{table} summarizes various cases that we have studied
numerically.  In addition to the $m_0=0$ cases discussed above, we
have also computed power-law decay rates for $m_0=1$ and
$m_0=2$. These cases allow us to test the predictions of
Eq.~(\ref{eq:naive}) and Eq.~(\ref{eq:hod}) more thoroughly.  For
example, initial data proportional to $Y_{21}$ is forbidden
to evolve by mode-mixing into any lower $Y_{\ell m}$ mode ($\ell=0$ is
forbidden by $m$ conservation and $\ell=1$ is forbidden by parity).
Therefore, the $t^{-(2l+3)}$ law predicts a tail decay rate of
$t^{-7}$, which is what we observe.

\subsection{Coordinate effects}
\label{sec:coordinate-effects}

It has been argued~\cite{Poisson2002} for a rotating weakly-curved
spacetime that the difference between the
predictions of Eq.~(\ref{eq:naive}) and Eq.~(\ref{eq:hod}) is related
to the choice of coordinates, and in particular, that
Eq.~(\ref{eq:hod}) is correct if the initial data were proportional
to a spherical
harmonic in spheroidal coordinates.
To test whether this might
be true for Kerr, we repeated
the evolution of initial data proportional to $Y_{40}$, but with a
small coordinate change:
We still take initial data of the
form~(\ref{initial}), but we choose
\begin{equation}
\Pi=\Pi_0(r_{\rm BL},\theta_{\rm BL},\phi),
\label{eq:initialdataBL}
\end{equation}
where $\theta_{\rm BL}$ is the Boyer-Lindquist coordinate defined by
$\cos \theta_{\rm BL} = z/r_{\rm BL}$. Because $\theta_{\rm BL} \ne
\theta$ and $r_{\rm BL} \ne r$, this initial data differs slightly
from the form~(\ref{initial}).  In fact, the magnitude of this
difference is only about one part in a thousand. 

For brevity, here and in the following we will refer to the evolution
of initial data~(\ref{eq:initialdataBL}) as the ``BL'' case, and we
will refer to the $\ell_0=4$ evolution shown in Figure~\ref{KerrRing}
as the ``KS'' case, since the two evolutions differ only in which
radial and polar coordinate (Boyer-Lindquist or Kerr-Schild) are used
in the expression~(\ref{initial}) for the initial data\footnote{
In principle, instead of evolving this BL initial data, we could
have expanded the data in terms of KS spherical harmonics and
suitable radial basis functions, and used linearity to compute the
result.  However, this would require knowledge of both the power-law
decay rate of each KS spherical harmonic and the mixing rates
between all pairs of KS spherical harmonics.}.
Note that the Kerr background is expressed in the same
coordinate system for both the BL and the KS evolutions, and
is given by Eqs.~(\ref{eq:KSg}--\ref{eq:KSK}). Note also
that the transformation between the BL and KS radial
and polar coordinates is exactly the 
transformation~(\ref{eq:spheroidaldef1}--\ref{eq:spheroidaldef2}) studied
in~\cite{Poisson2002}.

\begin{figure} 
\begin{center}
\includegraphics[width=3.0in]{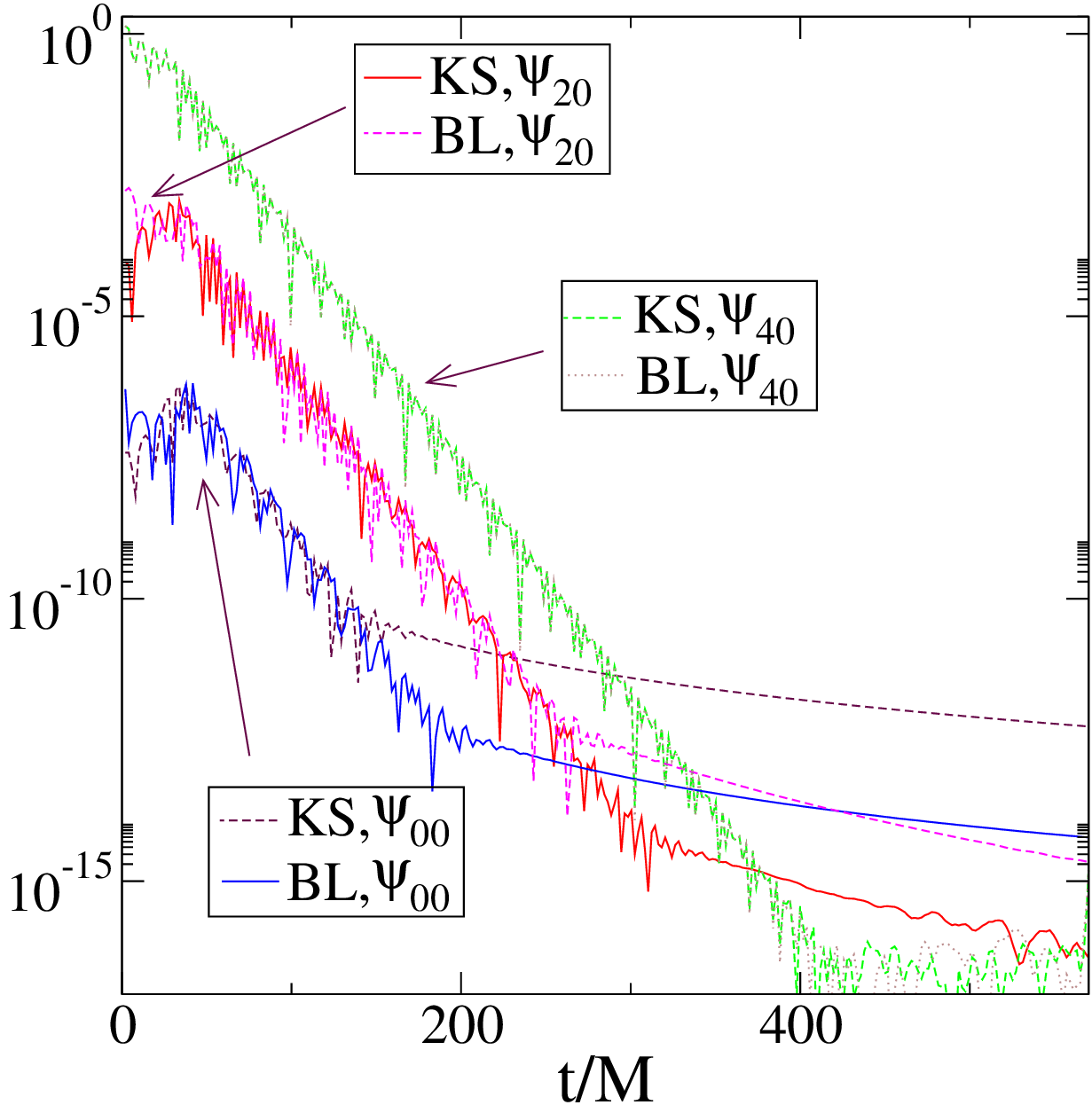}
\end{center}
\caption{Scalar field evolution in a Kerr ($a/M=0.5$) background
resulting from initial data proportional to $Y_{40}(\theta_{\rm
BL},\phi)$ (labeled by ``BL''), plus a corresponding evolution
resulting from initial data proportional to $Y_{40}(\theta,\phi)$ 
(labeled by ``KS''). Shown are the absolute values
of selected Kerr-Schild spectral coefficients
of $\psi$ for both cases. For both cases the resolution is 30/29.
The initial data for the KS and BL differ by only $0.1\%$, 
yet the details of the evolutions are quite different.}
\label{BLversusKS}
\end{figure}

\begin{figure} 
\begin{center}
\includegraphics[width=3.0in]{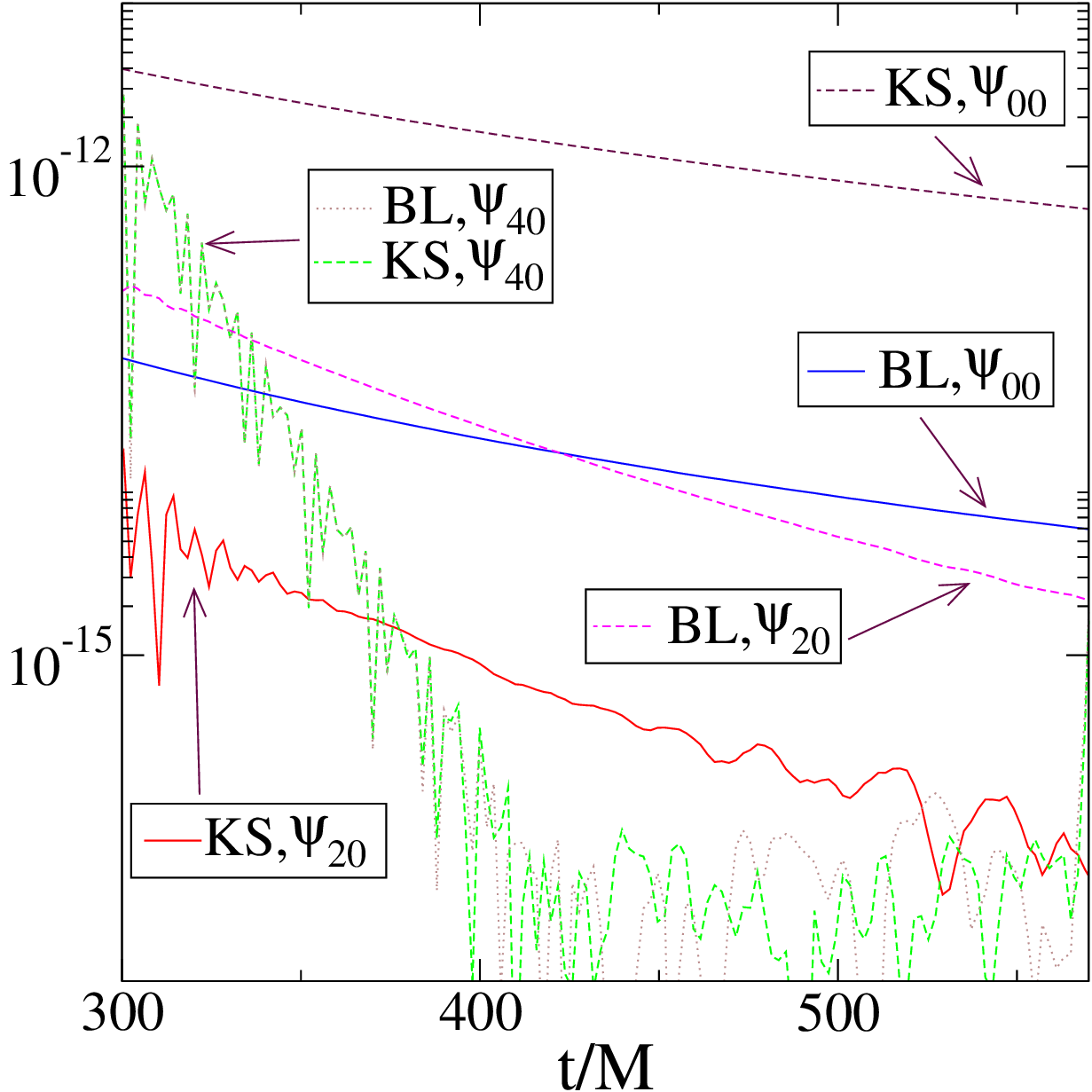}
\end{center}
\caption{Same as Fig.~\ref{BLversusKS} showing detail at late times.}
\label{BLversusKS2}
\end{figure}

If the argument of~\cite{Poisson2002} applies to the Kerr geometry,
then for the BL case, the power law falloff rate
should be $t^{-5}$, in agreement with Eq.~(\ref{eq:hod}), rather than
$t^{-3}$, which is predicted by Eq.~(\ref{eq:naive}).

It is quite difficult in the BL case to obtain an accurate value for
the late-time power index $\mu$ by the method used in
sections~\ref{sec:schwarzschild} and~\ref{sec:kerr-background}.  This
is because even though the initial data~(\ref{eq:initialdataBL})
differs only slightly from that used in the KS case, the evolution
proceeds quite differently: The resulting late-time tail is a few
orders of magnitude smaller than for the KS case, and by the time the
solution displays its late-time asymptotic behavior, the scalar field
time derivative is so small ($\sim 10^{-16}$) that machine roundoff
error (not numerical truncation error) obscures the results.

Fortunately, this roundoff error turns out to be largest at high angular
frequencies, so
it is still possible to determine the late-time behavior
of the BL case for low-frequency
spherical harmonic components of
the solution.  In Figures~\ref{BLversusKS} and~\ref{BLversusKS2},
different spherical harmonic components of the solution are plotted as
a function of time for both the BL and KS evolutions.  The spherical
harmonic components are computed by
\begin{equation}
\label{eq:sphericalharmonicintegral}
\psi_{\ell m} \equiv \int \psi Y_{\ell m}(\theta,\phi) d\Omega,
\end{equation}
where the integral is taken over the surface $r=11.75M$.  Note that
for all plots shown in Figures~\ref{BLversusKS} and~\ref{BLversusKS2},
the spherical harmonic appearing in the
integral~(\ref{eq:sphericalharmonicintegral}) is defined using the
Kerr-Schild $\theta$ and $\phi$ coordinates, and the integral is taken
over a surface of constant $r$, not a surface of constant $r_{\rm
BL}$.  Thus the quantities plotted in Figures~\ref{BLversusKS}
and~\ref{BLversusKS2} are in all cases the spherical harmonic
components of $\psi$ with respect to the Kerr-Schild coordinates.
Note also that $\psi_{\ell m}$ and the analogous quantity
$\dot{\psi}_{\ell m}$ can be used to compute a power index
for each individual spherical harmonic component.

For the KS evolution shown in Figures~\ref{BLversusKS}
and~\ref{BLversusKS2}, the initial data consist of pure $\ell=4$, but $\ell=2$
and $\ell=0$ components appear at very early times because of mode
mixing.  The tail of the $\ell=0$ component can be seen as early as
$t=150M$, but does not exceed the quasinormal ringing of the $\ell=4$
component until $t=300M$, after which it dominates.  Its measured
power index is $\mu=3.001 \pm 0.003$. The tail of the $\ell=2$
component can be seen for $t>300M$, but it is extremely small ($\sim
10^{-14}$). Its decay rate is roughly $t^{-7}$, but it is difficult to
determine the exponent accurately because it is buried in the
noise. The tail of the $\ell=4$ component cannot be seen; the $\ell=4$
component is buried in machine roundoff error after $t=400M$.

The BL evolution shown in Figures~\ref{BLversusKS}
and~\ref{BLversusKS2} is initially almost identical to the KS
case. Initially the BL case is not pure $\ell=4$ (recall $\ell$ here
refers to the index of the Kerr-Schild harmonic; the BL case {\it
is\/} pure $\ell=4$ with respect to Boyer-Lindquist spherical
harmonics), but also has a very small mixture of other components,
the largest being $\ell=6$ (not shown) and $\ell=2$.
As the BL case evolves in time, the tail of
the $\ell=0$ component first appears at $t=200M$, but it is a few
orders of magnitude smaller than the corresponding $\ell=0$ tail for
the KS case.  Its power index is $\mu=2.8\pm 0.3$. The tail of the
$\ell=2$ component, however, appears at $t=250M$ and is a few orders
of magnitude {\it larger\/} than the corresponding $\ell=2$ tail for
the KS case. Its power index is $\mu=7.0\pm 0.5$.  As in the KS case,
the tail of the $\ell=4$ component cannot be seen because of machine
roundoff error.

Although at intermediate times the $\ell=2$ mode is important for the
BL case, it is clear from Figure~\ref{BLversusKS2} that at very late
times, the $\ell=0$ mode will eventually dominate, resulting in a
decay rate of $t^{-3}$. In other words, the asymptotic decay rate
appears to be independent of whether the initial data are expressed in
terms of Kerr-Schild or Boyer-Lindquist spherical harmonics.  Thus,
the argument in~\cite{Poisson2002} apparently does not carry over to
the Kerr geometry. This is presumably because Kerr has strong-gravity
regions that influence the scalar field, and strong-gravity effects
were not included in~\cite{Poisson2002}.

We perform all our evolutions of BL initial data using a background
expressed in KS coordinates; this is because black hole excision
requires coordinates that are regular through the horizon.  A natural
question to ask is whether our results are different than they would
be if we had expressed our background in BL coordinates.  The answer
is yes, but only because we set ``BL initial data'' on a hypersurface
of constant KS time, not on a hypersurface of constant BL time.
Setting initial data on a hypersurface of constant BL time would
require an integration in BL coordinates (at least until the solution
were known on a full hypersurface of constant KS time, and from that
point on the evolution could proceed in KS coordinates), and is beyond
the scope of this paper.  However, for investigating the coordinate
effect described in~\cite{Poisson2002}, it is unnecessary to evolve
initial data on a BL time slice; the derivation of this effect
in~\cite{Poisson2002} involves no change in time slicing but only a
transformation of spatial coordinates, the same transformation that we
have done here.

It would also be interesting to repeat the evolutions in this section
with an outgoing initial pulse centered far from the black hole, so
that only the weak-gravity region is seen by the scalar field until
extremely late times. In this
case the weak-field approximation assumed by~\cite{Poisson2002} would
be valid for an extended period of time, so during this time
one should see a difference in decay rates between the BL
and KS evolutions.  Such a computation would be more difficult than
the ones presented here because it would require a more distant outer
boundary and therefore longer integration times. A single run
such as the one shown in Figures~\ref{BLversusKS} and~\ref{BLversusKS2} takes
about 23 hours on 30 processors of the IA-32 Linux cluster at NCSA;
the run time scales like $N_r^2$ if all subdomains have the same number
of radial collocation points $N_r$.
Future work may
involve a self-gravitating scalar field; in this case the equations
would be fully nonlinear.

\acknowledgments We thank Lee Lindblom and Richard Price for valuable
discussions.  Some computations were performed on the IA-32 Linux
cluster at NCSA. This research was supported in part by NSF grant
PHY-0099568 at the California Institute of Technology, NSF grant
PHY-9900672 at Cornell University, and NSF grants PHY-9734871 and 
PHY-0244605 at the
University of Utah.  ALE was supported by a LIGO Summer Undergraduate
Research Fellowship at the California Institute of Technology. LMB thanks
Kip Thorne for hospitality at Caltech during the beginning phase of
this work.


\end{document}